\documentclass[]{article}
\usepackage{amsmath}
\usepackage{graphicx}

\title{A Two Factor Forward Curve Model with Stochastic Volatility for Commodity Prices}
\author{Mark Higgins, PhD - Beacon Platform Incorporated}

\begin{document}

\maketitle

\begin{abstract}
	
We describe a model for evolving commodity forward prices that incorporates three important dynamics which appear in many commodity markets: mean reversion in spot prices and the resulting Samuelson effect on volatility term structure, decorrelation of moves in different points on the forward curve, and implied volatility skew and smile.

This model is a ``forward curve model" - it describes the stochastic evolution of forward prices - rather than a ``spot model" that models the evolution of the spot commodity price. Two Brownian motions drive moves across the forward curve, with a third Heston-like stochastic volatility process scaling instantaneous volatilities of all forward prices.

In addition to an efficient numerical scheme for calculating European vanilla and early-exercise option prices, we describe an algorithm for Monte Carlo-based pricing of more generic derivative payoffs which involves an efficient approximation for the risk neutral drift that avoids having to simulate drifts for every forward settlement date required for pricing.

\end{abstract}

\section{The Model Dynamics}

Forward price dynamics in commodities markets are often more complex than in traditional financial markets because the usual spot versus forward arbitrage cannot be implemented. The usual expression for a fair forward price in financial markets is

\begin{equation}
F(t,T) = S(t) e^{(R-Q) (T-t)}
\end{equation}

where $F(t,T)$ is the forward price for settlement at time $T$, as seen at calendar time $t$; $S(t)$ is the asset spot price at $t$, $R$ is the denominated currency discount rate, and $Q$ is the asset yield.

This expression holds only when two conditions are met: the asset can be stored, and the asset can be borrowed and shorted - both at a capacity comparable to the size of the market. This is the case for financial markets like equities, foreign exchange, and precious metals, but fails to hold for most physical commodity market such as oil, natural gas, and electricity.

If those conditions do hold then forward prices for different settlement dates tend to move with close to 100\% correlation, because any market forward price that deviated from the fair forward price would be arbitraged back into line. If these two conditions fail to hold, however, then forward prices can move relative to each other with less than a 100\% correlation since arbitrage does not force them to move together.

Models for commodity price evolution typically fall into two categories: models of the spot price and convenience yield $Q$ (for example \cite{schwartz} and \cite{JEHandJR}) and models of the forward prices themselves (for example \cite{gabillon} and \cite{zolotko}). Since commodity markets usually trade in forward price terms, through futures markets and over-the-counter forward transactions, the latter category of model - called ``forward curve" models - is often more intuitive for derivatives traders. Forward curve models are similar to interest rate models like the LIBOR Market Model \cite{rebonato}.

In addition to decorrelation of moves in forward prices, the absence of the two conditions required to implement the spot versus forward arbitrage also means that the risk neutral drift of the commodity spot price is not forced to be equal to the difference in two interest rates, and the risk neutral drift can have a more complicated structure. In particular, commodity markets exhibit mean reversion in spot prices \cite{geman}, and that real-world mean reversion can bleed into risk neutral pricing. Mean reversion in spot prices means that the instantaneous volatility of forward prices with longer settlement times tends to be smaller than instantaneous volatility of forward prices with shorter settlement times (the Samuelson effect, see \cite{samuelson}).

Decorrelation of forward price moves and mean reversion are two important dynamics to incorporate in any model of commodity forward price evolution. For example, the Gabillon model \cite{gabillon} is a two-factor forward curve model that incorporates both dynamics, and leads to lognormally-distributed forward prices with a term structure of implied volatility that can closely match at-the-money market option prices. A lognormal distribution of forward prices gives implied volatilities that are constant as a function of strike for a given expiration date, however, and real commodity markets exhibit significant implied volatility skew and smile.

The model defined in this paper adds one-factor stochastic volatility to the modeled dynamics, following a similar approach to the well known Heston stochastic volatility model, which generalizes a spot price model with stochastic volatility \cite{heston}. This generalization lets us model market implied volatility skew and smile in addition to the volatility term structure.

\section{The Model Definition}

Our model for evolution of a forward price $F(t,T)$ is given by the set of stochastic differential equations

\begin{eqnarray}
\frac{dF(t,T)}{F(t,T)} &=& \sqrt{v(t)} \sigma \left( e^{-\beta_1 (T-t)} dz_1(t) + R e^{-\beta_2 (T-t)} dz_2(t) \right) \nonumber \\
dv(t) &=& \beta (1 - v(t)) dt + \alpha \sqrt{v(t)} dz_3(t) \nonumber \\
E[dz_1(t) dz_2(t)] &=& \rho dt \nonumber \\
E[dz_1(t) dz_3(t)] &=& \rho_1 dt \nonumber \\
E[dz_2(t) dz_3(t)] &=& \rho_2 dt
\end{eqnarray}

That is: forward prices across the forward curve (all different values of $T$ for a given calendar time $t$) are driven by only two Brownian motions, and the impact of each Brownian motion shock is different for different values of $T$, leading to an instantaneous volatility that tends to decline with $T$ due to the constant mean reversions $\beta_1$ and $\beta_2$ that model the observed Samuelson effect.

The two forward curve shocks are correlated with a constant correlation $\rho$, and that lack of perfect correlation means that moves in forwards for different values of $T$ are not 100\%, allowing us to model the observed decorrelation of forward price moves.

Finally, the instantaneous volatility of all forward prices is scaled by the Heston factor $v(t)$, which is unitless and is order 1. We further specify the initial condition of $v(0) = 1$, so that volatility term structure is mostly specified through the constant model parameters $\sigma$, $\beta_1$, $\beta_2$, $R$, and $\rho$. The constant mean reversion speed of the volatility factor, $\beta$, and the constant volatility of volatility parameter $\alpha$, define the term structure of implied volatility smile. The correlation of volatility factor with first forward curve factor $\rho_1$, and with second forward curve factor $\rho_2$, define the term structure of implied volatility skew.

\section{Vanilla and Early Exercise Option Pricing}

``Vanilla" European options are ones where the expiration time $t_e$ of the option matches the settlement time of the underlying forward contract $T$. An ``early exercise" option is also a European option on a forward settling at time $T$, but the option expires at an earlier time $t_e \le T$, exercising into the underlying forward contract when in the money at $t_e$. Since vanilla options are a subset of early exercise options, in this section we derive pricing for early exercise options.

The approach is similar to that for pricing vanilla options under the original Heston model \cite{heston}: first we calculate the characteristic function of the log-forward price on the expiration time; then we integrate a function including that characteristic function to calculate an option price.

Unlike the standard Heston model, the characteristic function cannot be calculated in closed form: one numerical integration is required. Since we need another numerical integration to get an option price, this method imples a two-dimensional numerical integration to price early exercise options, rather than a one-dimensional numerical integration for the original Heston model. A two-dimensional numerical integration is still significantly more computationally efficient than other numerical techniques for pricing options, such as backward induction through a finite difference grid or Monte Carlo simulation. 

We start by changing variable to $x(t,T) = \ln \left(F(t,T)/F(0,T) \right)$. Then

\begin{equation}
dx = -\frac{v(t)}{2} \sigma^2_F(t,T) dt + \sqrt{v(t)} \sigma \left( e^{-\beta_1 (T-t)} dz_1(t) + R e^{-\beta_2 (T-t)} dz_2(t) \right)
\end{equation}

where $\sigma^2_F(t,T)$ is the determinstic part of the instantaneous volatility:

\begin{equation}
\sigma^2_F(t,T) = \sigma^2 \left( e^{-2 \beta_1 (T-t)} + R^2 e^{-2 \beta_2 (T-t)} + 2 \rho R e^{-(\beta_1 + \beta_2) (T-t)} \right)
\end{equation}

We define the characteristic function $f$ as

\begin{equation}
f(x(t,T), v(t); \theta) = E[e^{i \theta x(t_e,T)}]
\end{equation}

where the expectation $E[...]$ is under the risk neutral measure, $t$ is the current calendar time, $t_e$ is the expiration time of the early exercise option, and $T$ is the settlement date of the underlying forward.

Since we are considering only one settlement time $T$ in the pricing of early exercise options, going forward in this section we will drop the $T$ from $x(t,T)$ and write it simply as $x(t)$.

If we apply Ito's Lemma to $f$ and set its drift to zero (because $f$ is a martingale in the risk neutral measure) we get a partial differential equation that $f$ satisfies:

\begin{eqnarray}
\frac{v \sigma^2_F(t,T)}{2} (\frac{\partial^2 f}{\partial x^2} - \frac{\partial f}{\partial x}) + \beta (1-v) \frac{\partial f}{\partial v} + \frac{\alpha^2 v}{2} \frac{\partial^2 f}{\partial v^2}  \nonumber \\
+ \sigma \alpha v (e^{-\beta_1 (T-t)} \rho_1 + R e^{-\beta_2 (T-t)} \rho_2) \frac{\partial^2 f}{\partial x \partial v} + \frac{\partial f}{\partial t} = 0
\end{eqnarray}

The next step is to try a solution of the form

\begin{equation}
f(x(t), v(t); \theta) = e^{i \theta x(t) + A(t) + B(t) v(t)}
\end{equation}

where $A$ and $B$ are functions only of calendar time $t$. Also, we change the calendar time variable to $\tau = t_e - t$, so that the boundary condition at $\tau=0$ is $A = B = 0$.

In the usual way this reduces the partial differential equation to a pair of ordinary differential equations:

\begin{eqnarray}
\frac{dA}{d\tau} &=& \beta B \nonumber \\
\frac{dB}{d\tau} &=& -\frac{1}{2} (\theta^2 + i \theta) \sigma^2_F(t,T) - \beta B + \frac{\alpha^2}{2} B^2  \nonumber \\
&+& i \theta B \alpha \sigma (e^{-\beta_1 \tau} e^{-\beta_1 (T-t_e)} \rho_1 + R e^{-\beta_2 \tau} e^{-\beta_2 (T-t_e)} \rho_2)
\end{eqnarray}

Unlike the standard Heston model we know of no closed form solution to this pair of ordinary differential equations; instead they can be numerically integrated using standard numerical techniques for integrating first-order ordinary differential equations. Along with the initial conditions $A(\tau=0) = B(\tau=0) = 0$, we can calculate $A$ and $B$ for any $\tau$, and thereby calculate the characteristic function $f$.

The next step is to calculate a vanilla option price from the characteristic function. As usual, the call price $C(K)$ for strike $K$ is given by

\begin{equation}
C(K) = D(T) (F(0,T) - \frac{K}{2} - \frac{K}{\pi} \int_{\theta=0}^\infty {\Re[\frac{f(\theta) e^{-i \theta \ln K/F(0,T)}}{\theta^2 + i \theta}]})
\end{equation}

where $D(T)$ is the discount factor to settlement time $T$. Put prices can be calculated from call prices using put/call parity, of course, as these are European options.

\begin{figure}
	\includegraphics[width=\linewidth]{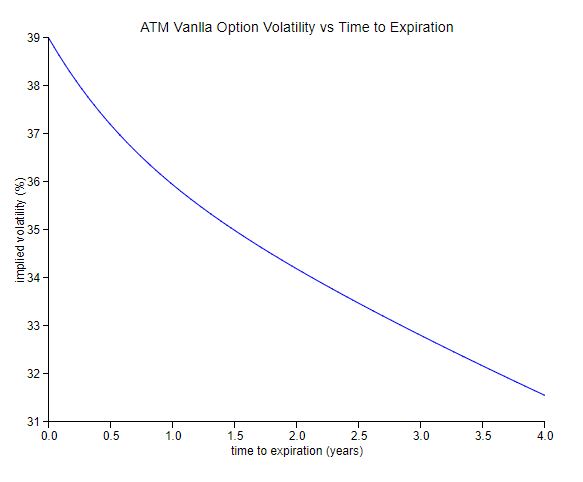}
	\caption{At-the-money forward implied volatility for vanilla options as a function of expiration time. These are vanilla options so $t_e = T$. The y-axis shows percentage implied volatility and the x-axis shows time to expiration in years.}
	\label{fig:atmvsexp}
\end{figure}

Figure \ref{fig:atmvsexp} shows at-the-money forward implied volatility as a function of time to expiration - the term structure of volatility - for vanilla options, where $t_e=T$. Note how the implied volatility decays with time to expiration because of the Samuelson effect, as represented in this model with the two mean reversion strengths $\beta_1$ and $\beta_2$. In this example the model parameters were $\sigma = 0.4$, $\beta_1 = 0.1$, $\beta_2 = 1$, $R = 0.5$, $\rho = -0.3$, $\beta = 0.5$, $\alpha=1$, $\rho_1 = 0.3$, and $\rho_2 = 0.3$. Note that in practice $\alpha \approx 1$ is a representative scale for the value of $\alpha$ in many commodity markets.

\begin{figure}
	\includegraphics[width=\linewidth]{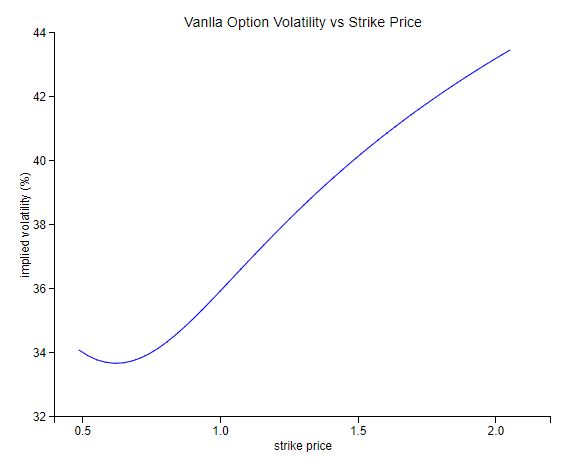}
	\caption{Vanilla implied volatility vs strike for a representative set of parameters. The y-axis shows percentage implied volatility and the x-axis shows option strike price (where the forward is equal to 1).}
	\label{fig:volvsstrike}
\end{figure}

Figure \ref{fig:volvsstrike} shows vanilla option implied volatility as a function of option strike price (with forward = 1) for a fixed time to expiration of $t_e=T=1$ (other parameters as in the previous example) which shows how the model creates implied volatility skew.

\section{Monte Carlo Simulation}

Many exotic derivatives depend on more than one forward price: that is, on $F(t,T)$ for a range of calendar times $t$ and settlement times $T$. For example, an average price option depends on daily forward price fixings during its life: one fixing for each business day (corresponding to $t$ values), and typically for the ``prompt" futures contract (nearest to settlement on the fixing date), which defines the settlement times $T$ for each fixing time $t$. That might mean dozens or even hundreds of separate forward curve dependencies.

In the lognormal two factor model, two separate stochastic factors can be simulated in a Monte Carlo simulation; given the values of those two factors at calendar time $t$, the forward price $F(t,T)$ can be calculated for any value of $T$. This dimensionality reduction (from the full set of market values of $T$ to the two stochastic factors) leads to effective numerical pricing of many derivatives.

The same is not true for this stochastic volatility extension. With $x(t,T) = \ln \frac{F(t,T)}{F(0,T)}$,

\begin{equation}
dx(t,T) = -\frac{1}{2} v(t) \sigma^2_F(t,T) dt + \sqrt{v(t)} \sigma \left( e^{-\beta_1 (T-t)} dz_1(t) + R e^{-\beta_2 (T-t)} dz_2(t) \right)
\end{equation}

If we define factors $u_1(t)$ and $u_2(t)$ as

\begin{equation}
du_i(t) = \sqrt{v(t)} e^{\beta_i t} dz_i(t)
\end{equation}

then we can write

\begin{equation}
dx(t,T) = -\frac{1}{2} v(t) \sigma^2_F(t,T) dt + \sigma \left( e^{-\beta_1 T} du_1(t) + R e^{-\beta_2 T} du_2(t) \right)
\end{equation}

As $x(0,T) = 0$ by construction we can integrate this to

\begin{equation}
x(t,T) = -\frac{1}{2} \int_{s=0}^t {v(s) \sigma^2_F(s,T) ds} + \sigma \left( e^{-\beta_1 T} u_1(t) + R e^{-\beta_2 T} u_2(t) \right)
\end{equation}

In the lognormal limit, where $v(t)=1$ always, the value of $x(t,T)$ is specified only by the two factor values $u_1(t)$ and $u_2(t)$. In the general case, however, the integrated drift term is no longer deterministic: it depends on the path of $v(t)$. Worse, that drift term depends on the settlement time $T$, so that a separate drift term is needed for each value of $T$.

If the payoff being priced depends on just one value of $T$ this is not a problem; that drift term can be calculated as an additional variable per Monte Carlo path, along with the two factors $u_1$ and $u_2$ and the Heston volatility factor $v$.

This is a computational problem for Monte Carlo simulation, however, if the derivative price depends on forwards for $N \gg 1$ values of $T$, because then we need to simulate $N$ separate drift terms on each path. For large $N$ that can become computationally inefficient or too memory intensive for practical use.

We solve this problem with an approximation for the drift term. Define $I(t,T)$ as

\begin{equation}
I(t,T) = \int_{s=0}^t {v(s) \sigma^2_F(s,T) ds} = \int_{s=0}^t {\sigma^2_F(s,T) ds} + \int_{s=0}^t {w(s) \sigma^2_F(s,T) ds}
\end{equation}

where $w(t) = v(t) - 1$. The first term is deterministic, so does not suffer the problem mentioned above. The second term, however, does suffer this problem, however, and here is where the approximation comes in: we approximate $I(t,T)$ as

\begin{equation}
I(t,T) \approx \int_{s=0}^t {\sigma^2_F(s,T) ds} + k(t,T) \int_{s=0}^t {w(s) ds}
\end{equation}

We choose $k(t,T)$ as a determinstic function such that the variance of $I(t,T)$ under the approximation matches the true variance. If this approximation is sufficiently accurate, we can run a Monte Carlo simulation and track just one extra path variable, the integrated Heston factor $\int_{s=0}^t {w(s) ds}$, and with that, the Heston volatility variable $v(t)$ and the two regular factors $u_1(t)$ and $u_2(t)$, we can calculate any forward price $F(t,T)$.

If we match variances we can write

\begin{equation}
Var(\int_{s=0}^t {w(s) \sigma^2_F(s,T) ds}) = Var(k(t,T) \int_{s=0}^t {w(s) ds})
\end{equation}

so that

\begin{equation}
k^2(t,T) = \frac{2 \int_{s_2=0}^t \int_{s_1=0}^{s_1} \sigma^2_F(s_1,T) \sigma^2_F(s_2,T) E[w(s_1) w(s_2)] ds_1 ds_2}{2 \int_{s_2=0}^t \int_{s_1=0}^{s_1} E[w(s_1) w(s_2)]}
\end{equation}

To calculate this we need to calculate $J(s_1,s_2) = E[w(s_1) w(s_2)]$ for $s_1 \le s_2$:

\begin{eqnarray}
J(s_1, s_2) &=& E[w(s_1) w(s_2)] = E[w(s_1)^2] + \int_{s=s_1}^{s_2} w(s_1) dw(s) \nonumber \\
            &=& E[w(s_1)^2] - \beta \int_{s_1}^{s_2} J(s_1, s) ds
\end{eqnarray}

Then

\begin{equation}
\frac{\partial J}{\partial s_2} = -\beta J(s_1, s_2)
\end{equation}

so

\begin{equation}
J(s_1, s_2) = c e^{-\beta (s_2 - s_1)}
\end{equation}

for some constant $c$. We know that $J(s_1, s_1) = E[w(s_1)^2]$, and from the standard Heston model we know

\begin{equation}
E[w(t)^2] = \frac{\alpha^2}{2 \beta} (1 - e^{-2 \beta t})
\end{equation}

so

\begin{equation}
J(s_1, s_2) = E[w(s_1) w(s_2)] = \frac{\alpha^2}{2 \beta} (1 - e^{-2 \beta s_1}) e^{-\beta (s_2 - s_1)}
\end{equation}

We can then evaluate the expression for this ``drift factor" $k(t,T)$ by evaluating the two integrals. The resulting form is quite complex but is still closed form, and is given in appendix A.

With this expression for the drift factor we can run Monte Carlo simulations for pricing many derivatives by simulating the two factors $u_1(t)$ and $u_2(t)$; the Heston volatility factor $v(t)$; and the integrated Heston factor $(\int_{s=0}^t {w(s) ds}$. With that information, at any point on a Monte Carlo path, we can calculate any forward price $F(t,T)$.

\section{Validating the Drift Approximation}

The approximation is exact in two limits:

\begin{itemize}
	\item Volatility of volatility is zero. In this case $w(t) = v(t) - 1$ is always zero and the second term in $I(t,T)$ above is always zero.
	\item $\sigma_F(t,T)$ is a constant - that is, if both $\beta_1$ and $\beta_2$ are zero, or if $\beta_1$ is zero and $R=0$.
\end{itemize}

The approximation is most extreme when there is significant volatility of volatility and significant term structure of $\sigma_F(t,T)$.

To test the size of approximation error in the case where it is expected to be largest, we first compare the expected value of the simulated forward between the factor-based Monte Carlo simulation above and an approximation-free simulation of the single forward price, using the same Monte Carlo paths and seed to minimize the difference in simulated forwards (the difference goes to zero in the two limits above). The forward is the best metric of approximation error since the approximation affects the risk neutral drift; if the approximation fails, the simulation will give an incorrect expected forward.

The test used parameters $t_e = 1$, $T = 2$, $\sigma = 0.6$, $\beta_1 = 0.01$, $\beta_2 = 1$, $R = 0.5$, $\rho = -0.3$, $\beta = 0$, $\rho_1 = 0.3$, and $\rho_2 = 0.3$. The market forward price was 1. We used 100 time steps and 100,000 paths in the Monte Carlo simulation. Values of $\alpha$ ranged from 0 to 3, which gave ATMF implied volatilities ranging from 57.4

\begin{figure}
	\includegraphics[width=\linewidth]{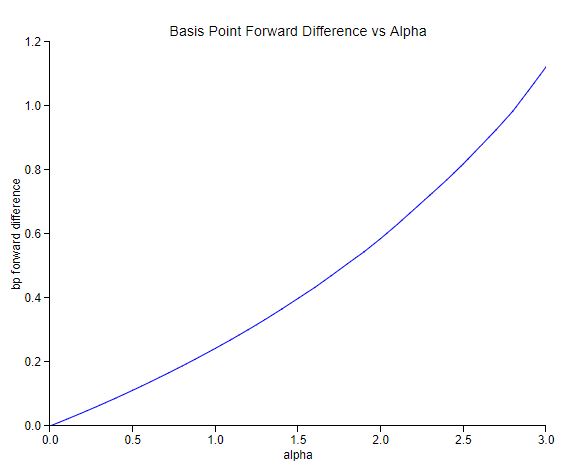}
	\caption{Approximation error in the simulated forward with extreme market and model parameters. The y-axis shows the basis point difference in the simulated forward, and the x-axis shows the value of $\alpha$.}
	\label{fig:simerror}
\end{figure}

\begin{figure}
	\includegraphics[width=\linewidth]{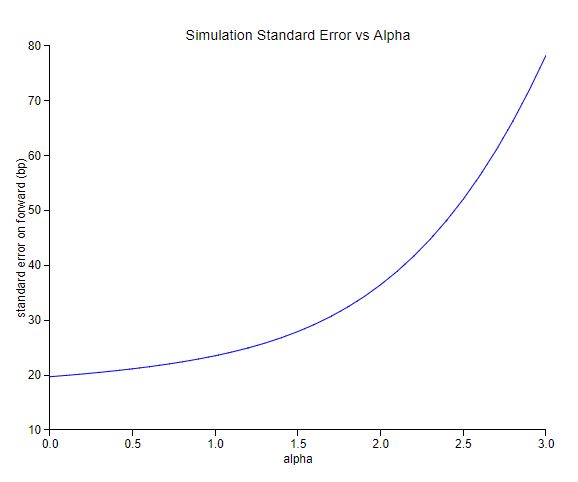}
	\caption{Standard error on the expected value of the forward in the Monte Carlo simulation with extreme market and model parameters. The y-axis shows the standard error in basis points, and the x-axis shows the value of $\alpha$.}
	\label{fig:stderr}
\end{figure}

Figure \ref{fig:simerror} shows the approximation error in basis points (one basis point is $10^{-4}$ of the forward) for these extreme parameters as a function of $\alpha$, the volatility of volatility, running from 0 to 3. Figure \ref{fig:stderr} shows the standard error on the simulated forward.

Note that the approximation error in the simulated forward is very small, reaching 1bp only in the most extreme market conditions where simulation error is similarly much larger and dominates the approximation error. For example, the standard error on the forward price with 100,000 paths for $\alpha=3$ is 78bp, 72 times larger than the approximation error.

\begin{figure}
	\includegraphics[width=\linewidth]{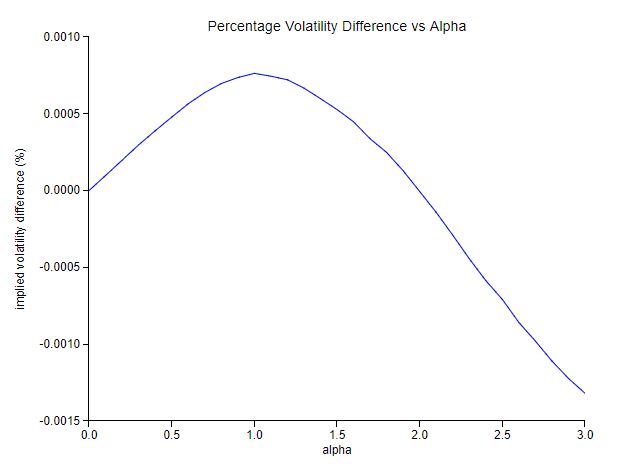}
	\caption{Approximation error in the simulated at-the-money implied volatility with extreme market and model parameters. The y-axis shows the implied volatility difference in percent, and the x-axis shows the value of $\alpha$.}
	\label{fig:volerror}
\end{figure}

\begin{figure}
	\includegraphics[width=\linewidth]{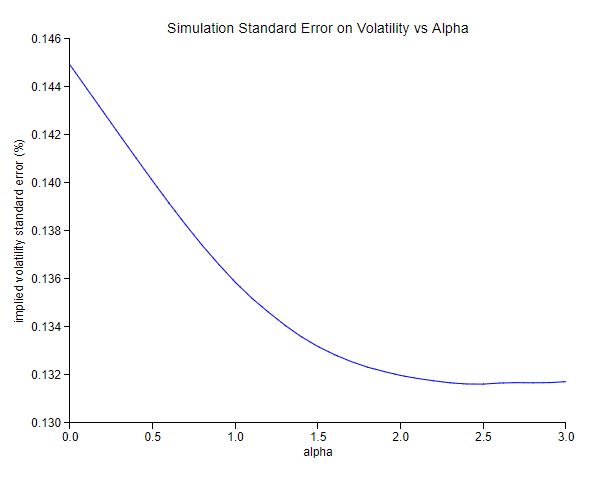}
	\caption{Standard error on the expected value of the at-the-money implied volatility in the Monte Carlo simulation with extreme market and model parameters. The y-axis shows the standard error in percentage volatility, and the x-axis shows the value of $\alpha$.}
	\label{fig:volstderr}
\end{figure}

The second test is for at-the-money implied volatility calculated under the two flavors of Monte Carlo simulation. Using the same parameters as in the first test, Figure \ref{fig:volerror} shows the approximation error in percentage volatility as a function of $\alpha$. Figure \ref{fig:volstderr} shows the standard error on the simulated implied volatility for the same range of $\alpha$.

Again, the approximation error is very small: maximally 0.0015\% in magnitude, versus a standard error from the simulation of around 0.14\%. The approximation error is insignificantly small for at-the-money implied volatility calculations, and hence vanilla option price calculations. The implied volatility level ranged from 30\% to 40\% across the range of $\alpha$ values with these parameters.

\begin{figure}
	\includegraphics[width=\linewidth]{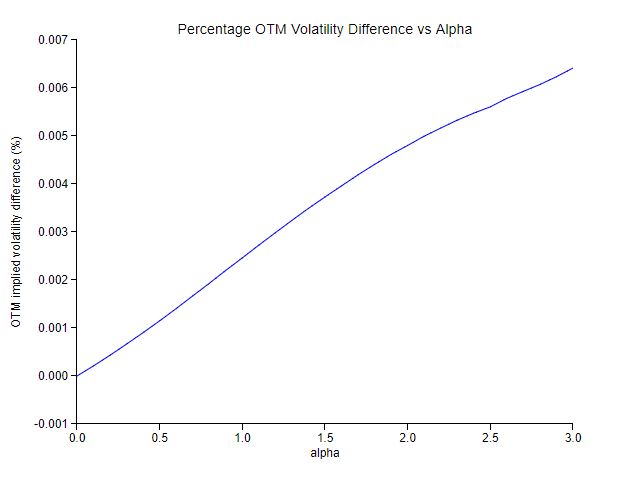}
	\caption{Approximation error in the simulated out-of-the-money implied volatility with extreme market and model parameters, for a strike 1.4 times the forward. The y-axis shows the implied volatility difference in percent, and the x-axis shows the value of $\alpha$.}
	\label{fig:otmerror}
\end{figure}

\begin{figure}
	\includegraphics[width=\linewidth]{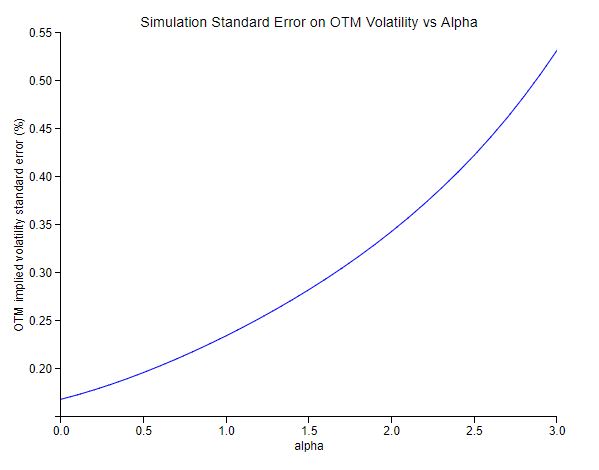}
	\caption{Standard error on the expected value of the out-of-the-money implied volatility in the Monte Carlo simulation with extreme market and model parameters, for a strike 1.4 times the forward. The y-axis shows the standard error in percentage volatility, and the x-axis shows the value of $\alpha$.}
	\label{fig:otmstderr}
\end{figure}

The final test is for out-of-the-money volatilities, where we use the same parameters as the at-the-money volatility test but use a strike of 1.4 times the forward - roughly one standard deviation out-of-the-money with a one-year expiration at roughly 40\% volatility. Figure \ref{fig:otmerror} shows the approximation error in percentage volatility as a function of $\alpha$. Figure \ref{fig:otmstderr} shows the standard error on the simulated implied volatility for the same range of $\alpha$.

As before, the approximation error is tiny - maximally 0.007\% - compared to the standard error on the implied volatility of 0.2-0.5\%.

The parameters used in these tests are relatively extreme, and approximation errors in more practical use cases will be even lower.

\section{Conclusions and Future Work}

We introduced a new model for the evolution of commodity forward prices in the context of pricing derivatives which includes mean reversion of commodity spot prices, decorrelation of moves across the forward curve, and implied volatility skew and smile. 

We included a recipe for efficiently calculating vanilla and early exercise option prices with a double numerical integration, which allows for relatively fast on-the-fly best fit of the model parameters to traded option prices. We also included an algorithm for pricing derivatives with a factor-based Monte Carlo simulation, under an approximation that avoids having to explicitly simulate integrated drift factors for forwards for every settlement date included in the payoff. The approximation error was validated as very small for practical cases.

Several areas for future work suggest themselves:

\begin{itemize}
	\item Adding deterministic time dependency to $\sigma$. As it stands, with constant model parameters, the model will not be able to exactly fit at-the-money implied volatility for all traded expirations. The usual approach for getting an exact fit is to let $\sigma$ be a piecewise-constant function of calendar time $t$ or settlement time $T$ (normally $\sigma(t)$ for non-seasonal markets like crude oil, and $\sigma(T)$ for seasonal markets like electricity and natural gas). 
	\item Adding deterministic time dependency to $\alpha$, $\rho_1$, and/or $\rho_2$. While constant values of $\alpha$, $\beta$, $\rho_1$, and $\rho_2$ can give rise to fairly generic smile and skew term structures, it is in general not possible to calibrate the model to reproduce market levels of implied volatility smile and skew for all traded expiration tenors with so few parameters. Letting the parameters be piecewise-constant functions of $t$ or $T$ allows for an exact calibration.
	\item Adding a second volatility factor. Some derivative prices are sensitive to correlation of moves of implied volatility across different expiration tenors, and that correlation is 100\% in this model because there is only one stochastic factor. Adding a second factor would give some control over that volatility correlation term structure.
\end{itemize}

\appendix

\section{Expression for Drift Factor}

The expression for the drift factor $k(t,T)$ used in the approximation of the risk neutral drift is a long but closed-form expression, shown here broken into pieces for ease of display:

\begin{equation}
k^2(t,T) = \frac{2 \sigma^4 \beta^2 e^{2 \beta t}}{e^{2 \beta t} (2 \beta t-3)+4 e^{\beta t}-1} \left(k_1(T) + k_2(t,T) + k_3(t,T) + k_4(t,T) + k_5(t,T) + k_6(t,T) \right)
\end{equation}

with

\begin{eqnarray}
k_1(T) &=& -\frac{2 \beta e^{-7 \beta_1 T-5 \beta_2 T} \left(k_{1a}(T)-k_{1b}(T) +k_{1c}(T)\right)
	\left(k_{1d}(T)- k_{1e}(T) +k_{1f}(T)\right)}{(\beta-2 \beta_1)^2 (\beta+2 \beta_1) (\beta-2 \beta_2)^2
	(-\beta+\beta_1+\beta_2)^2 (\beta+\beta_1+\beta_2) (\beta+2 \beta_2)} \nonumber \\
k_{1a}(T) &=&  \beta^2 \left(e^{2 (\beta_1-\beta_2) T} R^2+2 e^{(\beta_1-\beta_2) T} \rho  R+1 \right) \nonumber \\
k_{1b}(T) &=& \beta \left(\beta_2 \left(e^{2 (\beta_1-\beta_2) T} R^2+4
e^{(\beta_1-\beta_2) T} \rho  R+3\right)+\beta_1 \left(3 e^{2 (\beta_1-\beta_2) T} R^2+4
e^{(\beta_1-\beta_2) T} \rho  R+1\right) \right) \nonumber \\
k_{1c}(T) &=& 2 \left( \beta_2^2+\beta_1 \left(e^{2 (\beta_1-\beta_2) T}
R^2+4 e^{(\beta_1-\beta_2) T} \rho  R+1\right) \beta_2+\beta_1^2 e^{2 (\beta_1-\beta_2) T} R^2 \right) \nonumber \\
k_{1d}(T) &=&  \beta^4 \left(e^{5 \beta_1 T+3 \beta_2 T} R^2+2 e^{4 (\beta_1+\beta_2) T} \rho  R+e^{3 \beta_1 T+5 \beta_2 T} \right) \nonumber \\
k_{1e}(T) &=& e^{3 (\beta_1+\beta_2) T} \beta^2 \left(k_{1ea}(T) + k_{1eb}(T)+ k_{1ec}(T)\right) \nonumber \\
k_{1ea}(T) &=& \beta_1^2 \left(5 e^{2 \beta_1 T} R^2+8 e^{(\beta_1+\beta_2) T} \rho  R+e^{2
	\beta_2 T}\right) \nonumber \\
k_{1eb}(T) &=& 2 \beta_2 \left(e^{2 \beta_1 T} R^2+e^{2 \beta_2 T}\right) \beta_1 \nonumber \\
k_{1ec}(T) &=& \beta_2^2 \left(e^{2 \beta_1 T} R^2+8 e^{(\beta_1+\beta_2) T} \rho  R+5 e^{2 \beta_2 T}\right) \nonumber \\
k_{1f}(T) &=& 4 e^{3
	(\beta_1+\beta_2) T} \left(e^{2 \beta_1 T} R^2 \beta_1^4+2 \beta_2 e^{2 \beta_1 T} R^2 \beta_1^3+k_{1fa}(T) +2 \beta_2^3 e^{2
	\beta_2 T} \beta_1+\beta_2^4 e^{2 \beta_2 T}\right) \nonumber \\
k_{1fa}(T) &=& \beta_1^2 \beta_2^2 \left(e^{2 \beta_1 T} R^2+8 e^{(\beta_1+\beta_2) T} \rho  R+e^{2 \beta_2 T}\right)
\end{eqnarray}

\begin{eqnarray}
k_2(t,T) &=& \frac{2 \beta e^{-\beta t-7 (\beta_1+\beta_2) T}
	\left(k_{2a}(t,T)-k_{2b}(t,T)+k_{2c}(t,T)\right) \left(k_{2d}(t,T)-k_{2e}(t,T)+k_{2f}(t,T) \right)}{(\beta-2 \beta_1)^2 (\beta+2 \beta_1) (\beta-2 \beta_2)^2
	(-\beta+\beta_1+\beta_2)^2 (\beta+\beta_1+\beta_2) (\beta+2 \beta_2)} \nonumber \\
k_{2a}(t,T) &=& \left(e^{2 \beta_2 t+2 \beta_1 T} R^2+2 e^{(\beta_1+\beta_2) (t+T)} \rho  R+e^{2 \beta_1 t+2 \beta_2
	T}\right) \beta^2 \nonumber \\
k_{2b}(t,T) &=& \left(k_{2ba}(t,T)+k_{2bb}(t,T)\right) \beta \nonumber \\
k_{2ba}(t,T) &=& \beta_2 \left(e^{2 \beta_2 t+2 \beta_1 T} R^2+4 e^{(\beta_1+\beta_2) (t+T)} \rho  R+3 e^{2
	\beta_1 t+2 \beta_2 T}\right) \nonumber \\
k_{2bb}(t,T) &=& \beta_1 \left(3 e^{2 \beta_2 t+2 \beta_1 T} R^2+4 e^{(\beta_1+\beta_2) (t+T)}
\rho  R+e^{2 \beta_1 t+2 \beta_2 T}\right) \nonumber \\
k_{2c}(t,T) &=& 2 \left(e^{2 \beta_1 t+2 \beta_2 T} \beta_2^2+\beta_1
\left(e^{2 \beta_2 t+2 \beta_1 T} R^2+4 e^{(\beta_1+\beta_2) (t+T)} \rho  R+e^{2 \beta_1 t+2 \beta_2 T}\right)
\beta_2+\beta_1^2 e^{2 \beta_2 t+2 \beta_1 T} R^2\right) \nonumber \\
k_{2d}(t,T) &=& \left(e^{5 \beta_1 T+3 \beta_2 T} R^2+2 e^{4 (\beta_1+\beta_2) T} \rho  R+e^{3 \beta_1 T+5 \beta_2 T}\right) \beta^4 \nonumber \\
k_{2e}(t,T) &=& e^{3 (\beta_1+\beta_2) T}
\left(k_{2ea}(t,T)+k_{2eb}(t,T)+k_{2ec}(t,T)\right) \beta^2 \nonumber \\
k_{2ea}(t,T) &=& \left(5 e^{2 \beta_1 T} R^2+8 e^{(\beta_1+\beta_2) T} \rho  R+e^{2 \beta_2 T}\right) \beta_1^2 \nonumber \\
k_{2eb}(t,T) &=& 2 \beta_1 \beta_2 \left(e^{2 \beta_1 T} R^2+e^{2 \beta_2 T}\right) \nonumber \\
k_{2ec}(t,T) &=& \beta_2^2 \left(e^{2 \beta_1 T} R^2+8 e^{(\beta_1+\beta_2) T} \rho  R+5 e^{2 \beta_2 T}\right) \nonumber \\
k_{2f}(t,T) &=& 4 e^{3 (\beta_1+\beta_2) T} \left(e^{2
	\beta_1 T} R^2 \beta_1^4+2 \beta_2 e^{2 \beta_1 T} R^2 \beta_1^3+k_{2fa}(t,T)+2 \beta_2^3 e^{2 \beta_2 T}
\beta_1+\beta_2^4 e^{2 \beta_2 T}\right) \nonumber \\
k_{2fa}(t,T) &=& \beta_2^2 \left(e^{2 \beta_1 T} R^2+8 e^{(\beta_1+\beta_2) T} \rho  R+e^{2 \beta_2 T}\right) \beta_1^2
\end{eqnarray}

\begin{eqnarray}
k_3(t,T) &=& \frac{e^{(\beta_2-\beta_1) T}
	k_{3a}(t,T) k_{3b}(t,T)}{2 (\beta-2
	\beta_1)^2 (\beta-2 \beta_2)^2 (-\beta+\beta_1+\beta_2)^2 \left(R^2+2 e^{(\beta_2-\beta_1) T} \rho  R+e^{2
		(\beta_2-\beta_1) T}\right)} \nonumber \\
k_{3a}(t,T) &=& e^{(\beta_1-\beta_2) T} R^2+2 \rho  R+e^{(\beta_2-\beta_1) T} \nonumber \\
k_{3b}(t,T) &=& k_{3ba}(t,T)+k_{3bb}(t,T)+k_{3bc}(t,T)+k_{3bd}(t,T)+k_{3be}(t,T) \nonumber \\
k_{3ba}(t,T) &=& (\beta-2 \beta_1)^2 (-\beta+\beta_1+\beta_2)^2 e^{-4 \beta_2 T} R^4 \nonumber \\
k_{3bb}(t,T) &=& 4 (\beta-2 \beta_1)^2 (\beta-2 \beta_2) (\beta-\beta_1-\beta_2) e^{-(\beta_1+3 \beta_2) T} \rho  R^3 \nonumber \\
k_{3bc}(t,T) &=& 2 (\beta-2 \beta_1) (\beta-2 \beta_2) e^{-2 (\beta_1+\beta_2) T} \nonumber \\
& & \left(\left(2
\rho ^2+1\right) \beta^2 - 2 (\beta_1+\beta_2) \left(2 \rho ^2+1\right) \beta+\beta_1^2+\beta_2^2+2 \beta_1 \left(4
\beta_2 \rho ^2+\beta_2\right)\right) R^2 \nonumber \\
k_{3bd}(t,T) &=& 4 (\beta-2 \beta_1) (\beta-2 \beta_2)^2 (\beta-\beta_1-\beta_2) e^{-(3
	\beta_1+\beta_2) T} \rho  R \nonumber \\
k_{3be}(t,T) &=& (\beta-2 \beta_2)^2 (-\beta+\beta_1+\beta_2)^2 e^{-4 \beta_1 T}
\end{eqnarray}

\begin{eqnarray}
k_4(t,T) &=& -\frac{e^{-2 \beta t+3 \beta_1 t+\beta_2 t-\beta_1 T+\beta_2 T}
	k_{4a}(t,T) k_{4b}(t,T)}{2 (\beta-2 \beta_1)^2 (\beta-2 \beta_2)^2
	(-\beta+\beta_1+\beta_2)^2 \left(R^2+2 e^{(\beta_1-\beta_2) (t-T)} \rho  R+e^{2 (\beta_1-\beta_2)
		(t-T)}\right)} \nonumber \\
k_{4a}(t,T) &=& e^{(\beta_1-\beta_2) (T-t)} R^2+2 \rho  R+e^{(\beta_1-\beta_2) (t-T)} \nonumber \\
k_{4b}(t,T) &=& k_{4ba}(t,T)+k_{4bb}(t,T)+k_{4bc}(t,T) k_{4bd}(t,T) +k_{4be}(t,T)+k_{4bf}(t,T) \nonumber \\
k_{4ba}(t,T) &=&  (\beta-2 \beta_1)^2 (-\beta+\beta_1+\beta_2)^2 e^{-2 \beta_1 t+2 \beta_2 t-4 \beta_2 T} R^4 \nonumber \\
k_{4bb}(t,T) &=& 4 (\beta-2 \beta_1)^2 (\beta-2 \beta_2) (\beta-\beta_1-\beta_2) e^{\beta_2 (t-3 T)-\beta_1 (t+T)} \rho  R^3 \nonumber \\
k_{4bc}(t,T) &=& 2 (\beta-2 \beta_1) (\beta-2 \beta_2) e^{-2(\beta_1+\beta_2) T} \nonumber \\
k_{4bd}(t,T) &=& R^2 \left(\left(2 \rho ^2+1\right) \beta^2-2 (\beta_1+\beta_2) \left(2 \rho ^2+1\right) \beta+\beta_1^2+\beta_2^2 +2 \beta_1 \left(4 \beta_2 \rho ^2+\beta_2\right)\right) \nonumber \\
k_{4be}(t,T) &=& 4 (\beta-2 \beta_1) (\beta-2 \beta_2)^2 (\beta-\beta_1-\beta_2) e^{\beta_1 (t-3 T)-\beta_2 (t+T)} \rho R \nonumber \\
k_{4bf}(t,T) &=& (\beta-2 \beta_2)^2 (-\beta+\beta_1+\beta_2)^2 e^{2 \beta_1 t-2 \beta_2 t-4 \beta_1 T}
\end{eqnarray}

\begin{eqnarray}
k_5(t,T) &=& -\frac{k_{5a}(t,T) \left(k_{5b}(t,T)+k_{5c}(t,T)+k_{5d}(t,T)+k_{5e}(t,T)+k_{5f}(t,T)\right)}{k_{5g}(t,T) k_{5h}(t,T)} \nonumber \\
k_{5a}(t,T) &=& e^{(\beta_2-\beta_1) T} \left(e^{(\beta_1-\beta_2) T} R^2+2 \rho  R+e^{(\beta_2-\beta_1)T}\right) \nonumber \\
k_{5b}(t,T) &=& \beta_1 (\beta+2 \beta_1) (\beta_1+\beta_2) (\beta+\beta_1+\beta_2) (3 \beta_1+\beta_2) \nonumber \\
k_{5c}(t,T) &=& 8 \beta_1 (\beta+2 \beta_1) \beta_2 (\beta_1+\beta_2) (3
	\beta_1+\beta_2) (2 \beta+\beta_1+3 \beta_2) e^{-(\beta_1+3 \beta_2) T} \rho  R^3
	(\beta_1+3 \beta_2) e^{-4 \beta_2 T} R^4 \nonumber \\
k_{5d}(t,T) &=& k_{5da}(t,T) \left(\left(2 \rho ^2+1\right) \beta^2+2
	(\beta_1+\beta_2) \left(2 \rho ^2+1\right) \beta+\beta_1^2+\beta_2^2+2 \beta_1 \left(4 \beta_2 \rho^2
	+\beta_2\right)\right) R^2 \nonumber \\
k_{5da}(t,T) &=& 4 \beta_1 \beta_2 (3
	\beta_1+\beta_2) (\beta_1+3 \beta_2) e^{-2 (\beta_1+\beta_2) T} \nonumber \\
k_{5e}(t,T) &=& 8 \beta_1 \beta_2 (\beta_1+\beta_2) (2 \beta+3 \beta_1+\beta_2) (\beta+2 \beta_2)
	(\beta_1+3 \beta_2) e^{-(3 \beta_1+\beta_2) T} \rho  R \nonumber \\
k_{5f}(t,T) &=& \beta_2 (\beta_1+\beta_2) (\beta+\beta_1+\beta_2) (3
	\beta_1+\beta_2) (\beta+2 \beta_2) (\beta_1+3 \beta_2) e^{-4 \beta_1 T} \nonumber \\
k_{5g}(t,T) &=& 4 \beta_1 (\beta+2 \beta_1)
	\beta_2 (\beta_1+\beta_2) (\beta+\beta_1+\beta_2) (3 \beta_1+\beta_2) (\beta+2 \beta_2) (\beta_1+3 \beta_2) \nonumber \\
k_{5h}(t,T) &=& \left(R^2+2 e^{(\beta_2-\beta_1) T} \rho  R+e^{2 (\beta_2-\beta_1) T}\right)
\end{eqnarray}

\begin{eqnarray}
k_6(t,T) &=& \frac{k_{6a}(t,T)
	\left(k_{6b}(t,T)+k_{6c}(t,T)+k_{6d}(t,T) k_{6e}(t,T) +k_{6f}(t,T)+k_{6g}(t,T)\right)}{k_{6h}(t,T) k_{6i}(t,T)} \nonumber \\
k_{6a}(t,T) &=& e^{3 \beta_1 t+\beta_2
	t-\beta_1 T+\beta_2 T} \left(e^{(\beta_1-\beta_2) (T-t)} R^2+2 \rho  R+e^{(\beta_1-\beta_2) (t-T)}\right) \nonumber \\
k_{6b}(t,T) &=& \beta_1 (\beta+2 \beta_1) (\beta_1+\beta_2) (\beta+\beta_1+\beta_2) (3 \beta_1+\beta_2) (\beta_1+3
	\beta_2) e^{-2 \beta_1 t+2 \beta_2 t-4 \beta_2 T} R^4 \nonumber \\
k_{6c}(t,T) &=& 8 \beta_1 (\beta+2 \beta_1) \beta_2 (\beta_1+\beta_2)
	(3 \beta_1+\beta_2) (2 \beta+\beta_1+3 \beta_2) e^{\beta_2 (t-3 T)-\beta_1 (t+T)} \rho  R^3 \nonumber \\
k_{6d}(t,T) &=& 4 \beta_1 \beta_2 (3 \beta_1+\beta_2) (\beta_1+3 \beta_2) e^{-2 (\beta_1+\beta_2) T} \nonumber \\
k_{6e}(t,T) &=& R^2 \left(\left(2 \rho ^2+1\right) \beta^2+2(\beta_1+\beta_2) \left(2 \rho ^2+1\right) \beta+\beta_1^2+\beta_2^2 
	+ 2 \beta_1 \left(4 \beta_2 \rho \right)^2+\beta_2\right) \nonumber \\
k_{6f}(t,T) &=& 8 \beta_1 \beta_2 (\beta_1+\beta_2) (2 \beta+3 \beta_1+\beta_2) (\beta+2 \beta_2)
	(\beta_1+3 \beta_2) e^{\beta_1 (t-3 T)-\beta_2 (t+T)} \rho  R \nonumber \\
k_{6g}(t,T) &=& \beta_2 (\beta_1+\beta_2)
	(\beta+\beta_1+\beta_2) (3 \beta_1+\beta_2) (\beta+2 \beta_2) (\beta_1+3 \beta_2) e^{2 \beta_1 t-2 \beta_2 t-4\beta_1 T} \nonumber \\
k_{6h}(t,T) &=& 4 \beta_1 (\beta+2 \beta_1) \beta_2 (\beta_1+\beta_2) (\beta+\beta_1+\beta_2) (3
	\beta_1+\beta_2) (\beta+2 \beta_2) (\beta_1+3 \beta_2) \nonumber \\
k_{6i}(t,T) &=& R^2+2 e^{(\beta_1-\beta_2) (t-T)} \rho  R+e^{2(\beta_1-\beta_2) (t-T)}
\end{eqnarray}

\end{document}